# Hydrostatic pressure effect on $T_c$ of new $BiS_2$ based $Bi_4O_4S_3$ and $NdO_{0.5}F_{0.5}BiS_2$ layered superconductors

**G. Kalai Selvan[1], M. Kanagaraj[1], S. Esakki Muthu[1], Rajveer Jha[2], V. P. S. Awana[2] and S. Arumugam[*,1]**

[1]Centre for High Pressure Research, School of Physics, Bharathidasan University, Tiruchirappalli 620024, India
[2]Quantum Phenomena and Application Division, National Physical Laboratory (CSIR), Dr. K. S. Krishnan Road, New Delhi 110012, India

[*]Corresponding author: e-mail sarumugam1963@yahoo.com, Phone: - (O): 91-431- 2407118, Cell: 91-95009 10310; Fax No: - 91-431-2407045, 2407032.

We investigate the external hydrostatic pressure effect on the superconducting transition temperature ($T_c$) of new layered superconductors $Bi_4O_4S_3$ and $NdO_{0.5}F_{0.5}BiS_2$. Though the $T_c$ is found to have moderate decrease from 4.8 K to 4.3 K ($dT_c^{onset}/dP$ = -0.28 K/GPa) for $Bi_4O_4S_3$ superconductor, the same increases from 4.6 K to 5 K ($dT_c^{onset}/dP$ = 0.44 K/GPa) upto 1.31 GPa followed by a sudden decrease from 5 K to 4.7 K upto 1.75 GPa for $NdO_{0.5}F_{0.5}BiS_2$ superconductor. The variation of $T_c$ in these systems may be correlated to increase or decrease of the charge carriers in the density of states under externally applied pressure.

**1 Introduction** In the past 102 years many superconductors have been discovered and categorized into low temperature and high temperature superconductors. The low temperature superconductors which obey BCS theory had been well investigated in the past. In 1986, the first high $T_c$ of more than 30 K was found in $La_{2-x}Ba_xCuO_4$ by Bednorz and Muller [1]. Soon after, many new superconductors were discovered, and the maximum $T_c$ of 130 K was reported in $HgBa_2Ca_2Cu_3O_y$ at ambient pressure. Geo *et al.* found that on application of external pressure (31 GPa), the $T_c$ of $HgBa_2Ca_2Cu_3O_y$ enhanced dramatically to 164 K [2]. This study unveiled the importance of high pressure effect on high $T_c$ superconductors [2]. Recently, Kamihara *et al.*, reported that doping of $F^-$ atom in the place of $O^{2-}$ atom in LaFeAsO/F brought superconductivity with $T_c$ ~ 26 K [3]. These compounds are called Fe based pnictide superconductors [3], having quaternary ZrCuSiAs type tetragonal structure at room temperature, consisting of two alternating layers of insulating La-O and conducting Fe-As. Later new members of Fe based pnictide superconductors were reported with maximum $T_c$ of 56 K for $SmFeAsO_{1-x}F_x$ [4]. In addition to the chemical pressure (La/Sm substitution), application of external pressures also played an important and enhanced the $T_c$ of LaFeAsO/F from 26 K to 43 K ($dT_c/dP$= 4.25 K/GPa) [5]. Further, a maximum record of pressure dependence of $T_c$ (+9.1 K/GPa) has been reported in FeSe superconductors [6].

Hence, it is clear that a small change in the conducting layer by use of chemical doping of elements and/or externally applied pressure would modify the $T_c$ for the layered superconductors. Very recently, superconductivity with the $T_c$ of 4.4 K is found in $Bi_4O_4S_3$ polycrystalline compound [7]. It has the 2D tetragonal layered structure (space group I4/mmm) with stacking of $Bi_2S_4$, $Bi_2O_2$ and $SO_4$ layers. In this system, superconductivity is found in $BiS_2$ layer through the strong hybridization of Bi (6p) and S (3p) atoms. These compounds are called $BiS_2$ based superconductors. The doping of rare-earth elements (Ce, Pr, Yb) in the place of Ln in another $BiS_2$ based superconductor $LnO_{0.5}F_{0.5}BiS_2$ leads to increase of $T_c$ from 1.8 K to 5.3 K, [8,9]. A large shielding fraction of ~100 % and maximum onset of $T_c$ ~ 10 K has been reported in under high pressure (~2 GPa) synthesized $LaBiS_2O_{0.5}F_{0.5}$ [10]. Also, a new superconductor based on the rare earth blocking layer Nd-O in $NdO_{1-x}F_x$ $BiS_2$ ($0.1 \leq x \leq 0.6$) exhibited maximum $T_c$ of 6.7 K [11]. Kotegawa *et al.*, reported pressure effect on $T_c$ of $Bi_4O_4S_3$ superconductor and found that $T_c$ gradually decreases for all the applied pressures up to 1.92 GPa. However, the $T_c$ increases upto ~ 1GPa followed by decreasing trend upto 4 GPa in $LaBiS_2$(O/F) superconductor [12]. Keeping in view, the important role played by external pressure in discovery of new superconducting materials, we investigate the temperature dependence of electrical resistivity (300-3.5 K) measurements under various hydrostatic pressures for both $Bi_4O_4S_3$ (1.46 GPa) and $NdO_{0.5}F_{0.5}BiS_2$ (1.75 GPa) superconducting polycrystalline samples to understand the mechanism of superconducting transition temperature.

**2 Experiment** The polycrystalline samples of $Bi_4O_4S_3$ and $NdBiS_2O_{0.5}F_{0.5}$ were synthesized by two step solid state reaction method. The detailed investigation of synthesis and structural analysis were already described elsewhere [13,14]. The temperature dependence of electrical resistivity ($\rho$) at ambient and under various hydrostatic pressures is measured using four probe technique up to ~2 GPa using Closed Cycle Refrigerator – Variable Temperature Insert (CCR- VTI) system (REF-1810) in the temperature ranges of 300 to 3.5 K. A self-clamp type hybrid double cylinder (non-magnetic Ni-Cr-Al inner cylinder; Be-Cu outer cylinder) pressure cell was used to generate hydrostatic pressure up to 3.5 GPa. Daphne 7373 oil was used as a pressure transmitting medium. The applied pressure in the cell was estimated by the structural change of bismuth at room temperature.

**3 Results and discussion** Figs. 1(a & b) show the temperature dependence (300-3.5 K) of electrical resistivity ($\rho$) for superconducting $Bi_4O_4S_3$ and $NdO_{0.5}F_{0.5}BiS_2$ samples measured at ambient pressure. It is found that $\rho$ of both the compounds gradually decreases when the temperature reduces to 5 K, followed by sudden drop at $T_c$ onset of 4.8 K and 4.6 K corresponding to $Bi_4O_4S_3$ and $NdO_{0.5}F_{0.5}$ $BiS_2$ superconductor respectively. The observation of $T_c^{onset}$, $T_c^{mid}$ and $T_c^{offset}$ is indicated by arrows in the respective figures. The $T_c^{onset}$ is determined from the intersection of the two extrapolated lines, $T_c^{mid}$ is determined from the 50% of the drop in resistivity and $T_c^{offset}$ is determined as the resistivity reaches zero as shown in fig. 1(a & b).

The $\rho$ *vs* $T$ of $Bi_4O_4S_3$ and $NdO_{0.5}F_{0.5}BiS_2$ superconducting samples at various hydrostatic pressures are displayed in Figs. 2(a) and 2(b) respectively. It is clearly noted that the $T_c$ of $Bi_4O_4S_3$ decreases from 4.8 K to 4.3 K under applied pressure of upto 1.46 GPa. The insets of Figs. 2 (a & b) show $\rho(T)$ in the $T_c$ region (3.5 K to 6 K) on the corresponding samples to see the variation of $T_c$ under pressure clearly. Figs. 3(a& b) show $T_c$ *vs* $P$ (phase diagram) of $Bi_4O_4S_3$ and $NdO_{0.5}F_{0.5}BiS_2$ superconductors respectively. The negative pressure coefficient of – 0.287 K/GPa is observed in $Bi_4O_4S_3$, which is comparatively lower than the value (-1.1 K/GPa) reported by Kotegawa *et al.*, for the same compound. Such discrepancy between these two results might be connected with the inhomogeneity of pressure transmitting medium in the intender cell and piston cylinder cell. The data obtained from the piston cylinder setup is more reliable than the intender cell [12]. Under external pressure of 1.31 GPa, the $T_c^{onset}$ of $NdO_{0.5}F_{0.5}$ $BiS_2$ enhanced gradually from 4.6 K to 5 K, revealing a positive pressure coefficient of 0.4351 K/GPa. For Further higher pressures, the $T_c^{onset}$ decreased from 5 K to 4.7 K ($dT_c^{onset}/dP$= -0.1714 K/GPa) as the pressure increased from 1.31 GPa to 1.75 GPa, as shown in Fig. 3(b).

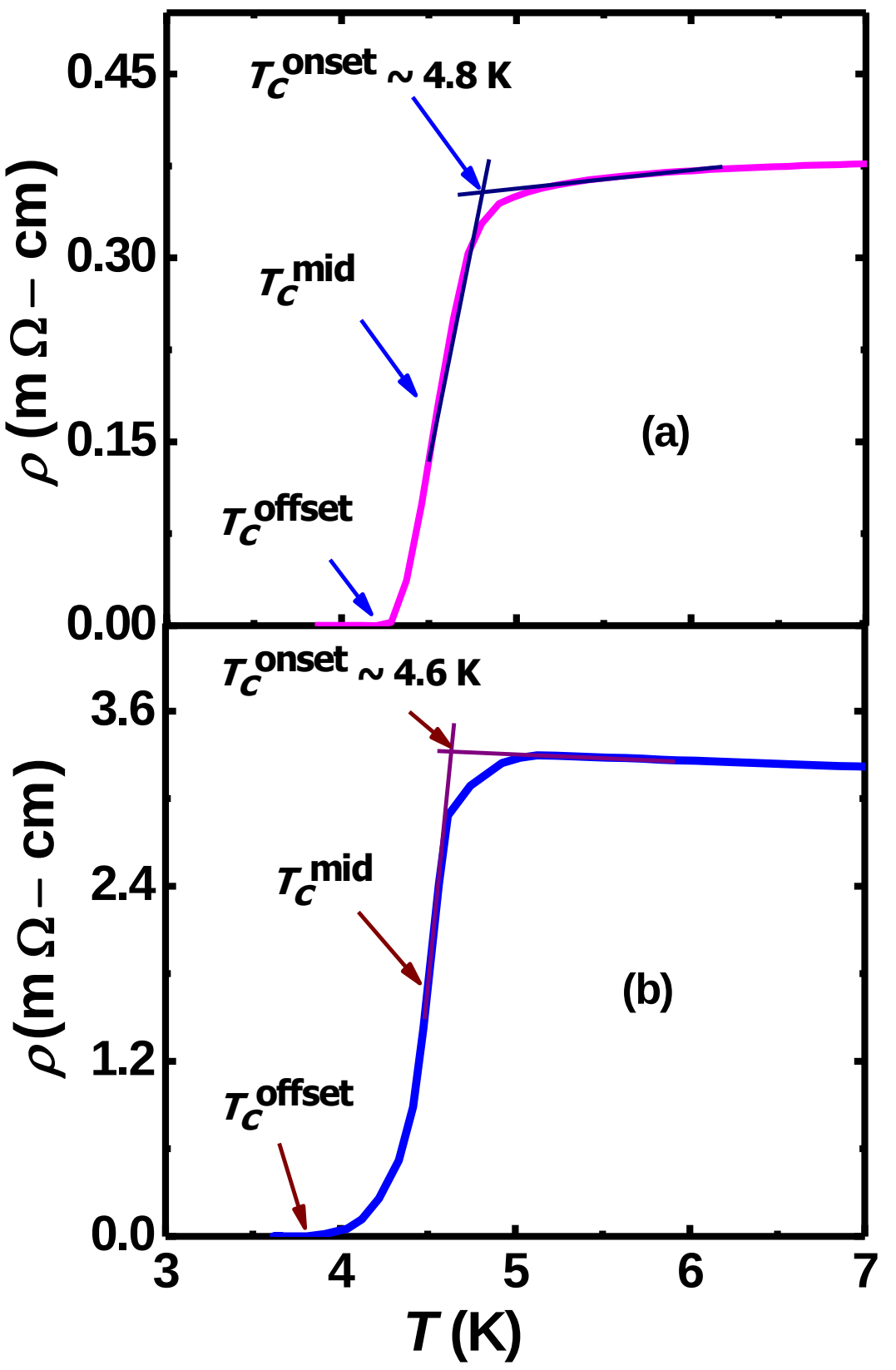

**Figure 1** Temperature dependent electrical resistivity of (a) $Bi_4O_4S_3$ and (b) $NdO_{0.5}F_{0.5}BiS_2$ at ambient pressure. The methodology of $T_c$ marking is also indicated in the same figures.

Hence it is clear that pressure dependence of $T_c$ and pressure coefficient in $Bi_4O_4S_3$ and $NdO_{0.5}F_{0.5}BiS_2$ superconductors display different trends. The present investigation of the pressure dependence of $T_c$ in $NdO_{0.5}F_{0.5}BiS_2$ superconductor resembles to pressure dependence of FeAs based 1111 type pnictide superconductors [13]. In our case, the polycrystalline $Bi_4O_4S_3$ compound shows the reduction of $T_c$ for all the applied pressures (0 to 1.46 GPa) and it may be due to existence of impurity phases (Bi, $Bi_2S_3$), which has been reported previously [14, 15]. Kotegawa *et al.* suggested that the $T_c$ of these new superconductors ($Bi_4O_4S_3$ and $LaBiS_2O_{0.5}F_{0.5}$) have been located between boundary of semiconducting/metallic region and it could be modified with respect to externally applied pressure.

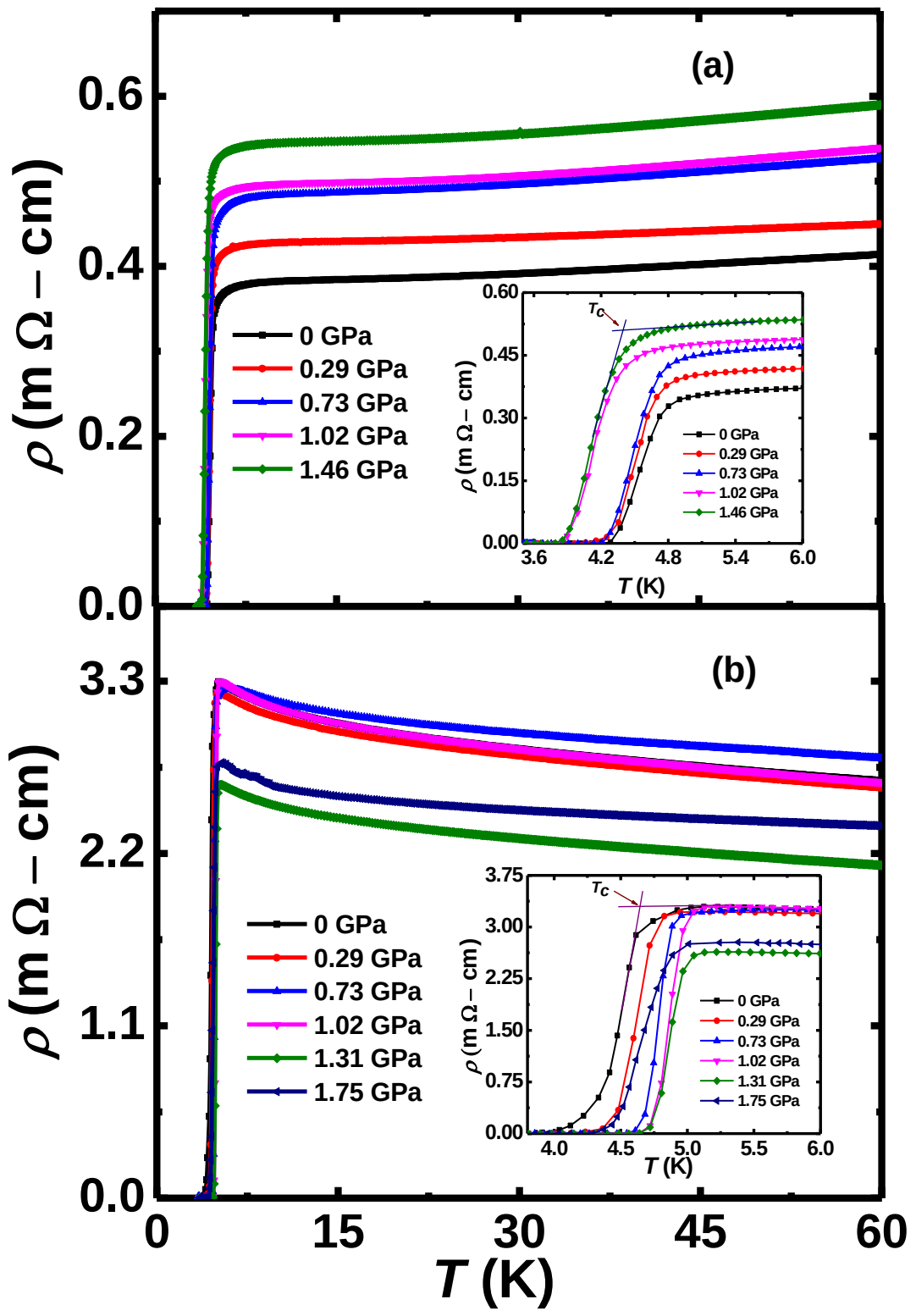

**Figure 2** The normal state resistivity ($\rho$) as a function of temperature *(T)* at various external hydrostatic pressures for (a) $Bi_4O_4S_3$ and (b) $NdO_{0.5}F_{0.5}BiS_2$. Insets of both figures denoted an enlarged view of $T_c$ variation in the low temperature region under various pressures.

The temperature dependence of electrical resistivity shows semiconducting nature before it becomes superconductor at ambient pressure, and slowly changed into metallic nature at a pressure of ~3.4 GPa. Further, pressure dependence of $T_c$ shows increasing trend upto 1.16 Gpa followed by decreasing trend for further increase of pressure upto ~4 GPa. Summarily, our results on temperature dependence of resistivity at various pressures and pressure dependence of $T_c$ show a similar trend upto 1.75 GPa, and consistence with the results reported by Kotegawa *et al.*

At high pressure region above 1.31 GPa both compounds found to display similar effect on $T_c$ under pressure, see Figs. 3(a)&(b). Obviously, the superconductor with positive pressure coefficient denotes an increase of density of charge carriers at the Fermi surface, whereas negative pressure coefficient on $T_c$ meant a reduction of charge carriers at density of states (DOS) [16]. At high pressure of 1.75 GPa the magnitude of the normal state resistivity increases and maintains the semiconducting nature before it becomes superconductivity in $NdO_{0.5}F_{0.5}BiS_2$ superconductor, see Fig. 2(b). This may be accounted for an unstable Fermi surface with respect to rate of change of carrier con**centration between conducting ($BiS_2$) and insulating** (NdO/F) layers under pressure.

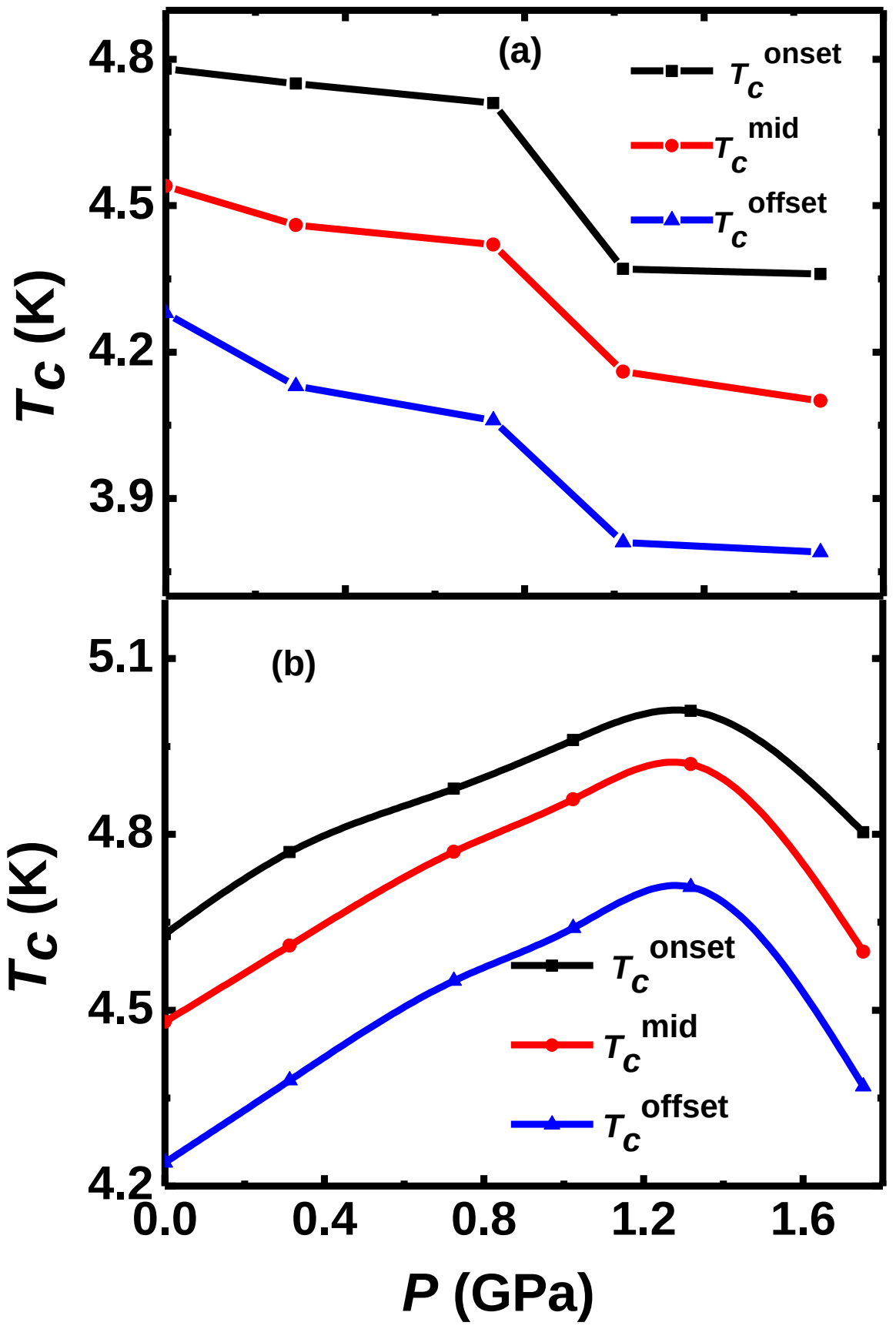

**Figure 3** $T_c^{onset}$, $T_c^{mid}$ and $T_c^{offset}$ vs applied pressure phase diagram for (a) $Bi_4O_4S_3$ and (b) $NdO_{0.5}F_{0.5}BiS_2$.

To be honest, we are unable to judge the exact reasons for the pressure dependence of $T_c$ and mechanism of superconductivity in $BiS_2$ based novel superconductors from our transport measurements under pressure. The origin of such intellectual arguments arises due to less number of high pressure measurements and review on $BiS_2$ based superconducting materials. Though, to the best of our knowledge, some other possible reasons may be responsible for the reduction of $T_c$ at high pressure: 1) a structural distortions 2) collapsed volume phase [17], 3) a strong hybridization of rare earth metal $Nd^{3+}$ (4f) and post transition metal $Bi^{3+}$(6P) with localized electrons resulting in a scattering of magnetic impurities 4) induced Kondo effect at low temperature [18]. The detailed theoretical work on DOS at high pressure and structural analysis by high pressure X-ray diffraction and neutron diffraction could be useful for better understanding of the change of $T_c$ at high pressure, and mechanism of $T_c$ of these superconducting materials.

In summary, we have carried out the temperature dependence high pressure electrical resistivity measurements up to ~1.8 GPa for $BiS_2$ based $Bi_4O_4S_3$ and $NdO_{0.5}F_{0.5}BiS_2$ novel layered superconductors. Also, we have investigated the variation of $T_c$ at different applied pressures using four probe techniques. The $T_c^{onset}$ decreases gradually from 4.8 K to 4.3 K in $Bi_4O_4S_3$ for all the applied pressures (0 to 1.4 GPa) and therefore it holds a negative pressure coefficient of -0.287 K/GPa. Contrarily, the pressure effect on $T_c$ of $NdO_{0.5}F_{0.5}BiS_2$ has positive pressure coefficient where the $T_c^{onset}$ increases up to 1.31 GPa at a rate of 0.435 K/GPa and thereafter it reduces for further enhancement of pressure becomes negative coefficient ($dT_c^{onset}/dP$= -0.1714 K/GPa K/GPa) on $T_c$ at high pressure of 1.75 GPa. The present results of hydrostatic pressure studies on electrical resistivity of these two $BiS_2$ based superconductors from $\rho$ *vs* $T$ studies suggest that the change of $T_c$ and superconducting state under pressure obtained for both compounds are to quite different from each other. Further theoretical and other experimental studies are required to understand the $BiS_2$ based series of superconducting materials.

**Acknowledgements** The author S. Arumugam wishes to thank DST (SERC, IDP, FIST) and UGC for the financial support. G. Kalai Selvan would like to thank UGC- BSR for the meritorious fellowship. M. Kanagaraj and S. Esakki Muthu, acknowledge the CSIR for the fellowship. Rajveer Jha acknowledges the CSIR for the senior research fellowship. This research at NPL is supported by DAE-SRC outstanding investigator scheme to work on search for new superconductors.